\documentclass[aps,pra,reprint,letterpaper,groupedaddress,floatfix,superscriptaddress]{revtex4-2}

\usepackage{multibib} 
\newcites{sup}{Supplementary References}
\usepackage{docmute} 
\usepackage{float}    

\usepackage{amsmath}
\usepackage[
colorlinks=true,
linkcolor=blue,
filecolor=blue,
citecolor=blue,
urlcolor=blue,
]{hyperref}
\usepackage{xurl}
\usepackage{graphicx}
\usepackage{color}
\usepackage{xcolor}

\makeatletter
\let\MYcaption\@makecaption
\makeatother
\usepackage[singlelinecheck=false, justification=raggedright, labelformat=simple, hypcap=true]{subcaption}

\makeatletter
\let\@makecaption\MYcaption
\makeatother

\usepackage{siunitx}
\usepackage{braket}
\usepackage{comment}
\usepackage{dcolumn}
\usepackage{threeparttable}
\DeclareSIUnit{\cps}{cps}
\usepackage{tabularx}
\usepackage{booktabs}
\graphicspath{{./}}

\newcommand{\setupSupplementary}{
    \clearpage
    \onecolumngrid 

    \setcounter{figure}{0}
    \setcounter{table}{0}
    \setcounter{equation}{0}
    \setcounter{section}{0}

    \renewcommand{\thefigure}{S\arabic{figure}}
    \renewcommand{\thetable}{S\arabic{table}}
    \renewcommand{\theequation}{S\arabic{equation}}
    \renewcommand{\thesection}{\arabic{section}}

    \renewcommand{\theHfigure}{S\arabic{figure}}
    \renewcommand{\theHtable}{S\arabic{table}}
    \renewcommand{\theHequation}{S\arabic{equation}}
    \renewcommand{\theHsection}{S\arabic{section}}

}

\makeatletter
\providecommand{\href@noop}[0]{\@secondoftwo}
\makeatother

\begin{document}

\title{Environment-Assisted Decoherence Suppression of Optical Non-Gaussian States}

\author{Akihiro Machinaga}
\affiliation{
    Department of Applied Physics, School of Engineering, The University of Tokyo,
    7-3-1 Hongo, Bunkyo-ku, Tokyo 113-8656, Japan
}
\author{Naoki Aritomi}
\affiliation{
    Department of Applied Physics, School of Engineering, The University of Tokyo,
    7-3-1 Hongo, Bunkyo-ku, Tokyo 113-8656, Japan
}
\author{Ryoga Sakurada}
\affiliation{
    Department of Applied Physics, School of Engineering, The University of Tokyo,
    7-3-1 Hongo, Bunkyo-ku, Tokyo 113-8656, Japan
}
\author{Daichi Okuno}
\affiliation{
    Department of Applied Physics, School of Engineering, The University of Tokyo,
    7-3-1 Hongo, Bunkyo-ku, Tokyo 113-8656, Japan
}
\author{Keitaro Anai}
\affiliation{
    Department of Applied Physics, School of Engineering, The University of Tokyo,
    7-3-1 Hongo, Bunkyo-ku, Tokyo 113-8656, Japan
}

\author{Takahiro Kashiwazaki}
\affiliation{
    Device Technology Labs, NTT, Inc., 3-1, Morinosato Wakamiya, Atsugi, Kanagawa 243-0198, Japan
}
\author{Takeshi Umeki}
\affiliation{
    Device Technology Labs, NTT, Inc., 3-1, Morinosato Wakamiya, Atsugi, Kanagawa 243-0198, Japan
}
\author{Shigehito Miki}
\affiliation{
    Advanced ICT Research Institute, National Institute of Information and Communications Technology, 
    588-2 Iwaoka, Nishi, Kobe, Hyogo 651-2492, Japan
}

\author{Masahiro Yabuno}
\affiliation{
    Advanced ICT Research Institute, National Institute of Information and Communications Technology, 
    588-2 Iwaoka, Nishi, Kobe, Hyogo 651-2492, Japan
}
\author{Hirotaka Terai}
\affiliation{
    Advanced ICT Research Institute, National Institute of Information and Communications Technology, 
    588-2 Iwaoka, Nishi, Kobe, Hyogo 651-2492, Japan
}
\author{Petr Marek}
\affiliation{
    Department of Optics, Palacky University, 17. listopadu 1192/12, 77146 Olomouc, Czech Republic
}
\author{Radim Filip}
\affiliation{
    Department of Optics, Palacky University, 17. listopadu 1192/12, 77146 Olomouc, Czech Republic
}
\author{Shuntaro Takeda}
\email{takeda@ap.t.u-tokyo.ac.jp}
\affiliation{
    Department of Applied Physics, School of Engineering, The University of Tokyo,
    7-3-1 Hongo, Bunkyo-ku, Tokyo 113-8656, Japan
}

\begin{abstract}
    Optical loss is a common bottleneck in photonic quantum information processing, undermining the quantum advantage over classical approaches. Although several countermeasures, such as quantum distillation and error correction, have been proposed, they typically require experimentally demanding non-Gaussian operations. Here, we demonstrate a Gaussian-only scheme that suppresses loss-induced decoherence for general, unknown optical quantum states. By injecting a squeezed vacuum state into an environment of the loss channel and performing feedforward based on environmental monitoring, the scheme effectively suppresses loss-induced noise. Our programmable loop-based optical circuit allows us to implement the scheme for several types of loss-sensitive non-Gaussian states under various loss conditions for up to five steps, and directly compare the results with the \textit{unsuppressed} case. Our results show that the scheme consistently mitigates state degradation, preserving higher fidelity and Wigner negativity than without suppression. This approach can be applied to mitigating a broad class of errors in optical systems and extending quantum memory lifetimes. Moreover, it is compatible with other loss-suppression techniques and extendable to physical platforms beyond optics, offering a promising route toward reducing the overhead required for fault-tolerant quantum information processing.
\end{abstract}
\maketitle

\section*{Introduction}

Quantum information processing using light has broad
applications in quantum computing, quantum communication,
and quantum sensing~\cite{flamini2019photonic}.
However, decoherence arising from imperfect optical components and environmental fluctuations still severely limits the practical implementation of these applications. One of the dominant sources of decoherence common to all these applications is optical loss.
Loss degrades the quantum nature of optical quantum states, eliminating
the quantum advantage in quantum information processing. For example, if loss degrades the
quality of squeezed light, Gaussian boson sampling using
it becomes classically simulatable~\cite{oh2024classical}, and the sensitivity enhancement in quantum sensing~\cite{lawrie2019quantum} is also degraded. Moreover, the negativity of
the Wigner function, essential for quantum computational
advantage~\cite{mari2012positive}, can easily vanish due to loss. In quantum communication, the quantum capacity vanishes when the transmission loss exceeds 50\%~\cite{wolf2007quantum}.

Various methodologies have been proposed to protect quantum states of light from the effects of loss. 
Representative approaches include quantum distillation~\cite{eisert2004distillation, takahashi2010entanglement} and quantum error correction~\cite{chuang1997bosonic, gottesman2001encoding}. However, these require resources or processing involving some form of nonlinearity (non-Gaussian elements) and thus carry high implementation costs~\cite{Niset2009, Eisert2002, Giedke2002, Fiurek2002}.
In contrast, several alternative methods relying solely on linear (Gaussian) optical operations have been proposed to mitigate the effects of loss under specific constraints, enabling easier implementation~\cite{serafini2004minimum, filip2003continuous, Marek2004, zhang2018quantum, provaznik2025adapting}. For example, a method has been realized that slows down the loss decoherence by preprocessing a known state with a negative Wigner function before it undergoes loss~\cite{LeJeannic2018, pan2023protecting}. Alternatively, approaches focusing on accessing the environment rather than the state have been realized, including a technique to remove loss effects from measurements of coherent states~\cite{Sabuncu2010}. 
However, no method has been realized that is applicable to general and unknown quantum states.

Here, we demonstrate a method to suppress loss-induced decoherence that works for general, unknown optical quantum states. Based on the previous proposal~\cite{Marek2004, zhang2018quantum}, we implement an environment-assisted decoherence suppression (EADS) scheme, where environmental degrees of freedom are controlled and monitored to mitigate noise. Specifically, our approach suppresses decoherence by preparing the environment in a specific quantum state—a squeezed vacuum state—and subsequently monitoring the environmental modes to perform feedforward operations to cancel the noise. A key advantage of this scheme is that, unlike conventional error correction protocols that typically require non-Gaussian elements, our implementation achieves decoherence suppression entirely within the Gaussian regime. 
To demonstrate its effectiveness across a broad range of scenarios, we verify the protocol using several types of loss-sensitive non-Gaussian states under diverse loss conditions and varying numbers of repetition steps up to five, alongside the \textit{unsuppressed} cases for comparison. Such systematic verification is made possible by employing a programmable loop-based optical circuit~\cite{Yoshida2025}. 
Our results demonstrate that this scheme consistently mitigates the degradation of the states, maintaining higher fidelity and more pronounced quantum non-Gaussian features compared to the case of \textit{unsuppressed} decoherence across various conditions. 

Our work provides a versatile strategy for mitigating loss-induced decoherence in optical systems, including losses arising from imperfect optical components, mode mismatch, mode crosstalk, and imperfect quantum transduction~\cite{richardson2013space, slussarenko2019photonic, zhang2018quantum, werner2026electro}. The approach enables extended lifetimes of quantum memories in the presence of loss~\cite{Marek2004, okuno2024time, kaneda2015time, simon2024experimental} and is broadly applicable to other continuous-variable platforms, such as superconducting microwave circuits~\cite{campagne2020quantum, eickbusch2022fast} and trapped-ion vibrational modes~\cite{fluhmann2019encoding}. Its compatibility with existing error-correction and loss-suppression schemes~\cite{chuang1997bosonic, gottesman2001encoding, eisert2004distillation, takahashi2010entanglement, LeJeannic2018} further provides a promising path toward reducing the overhead for fault-tolerant quantum information processing~\cite{fukui2018high, guillaud2019repetition}.

\section{Concept of environment-assisted decoherence suppression\label{sec:method}}
\begin{figure*}[t]
    \centering
    \includegraphics[width=0.80\linewidth]{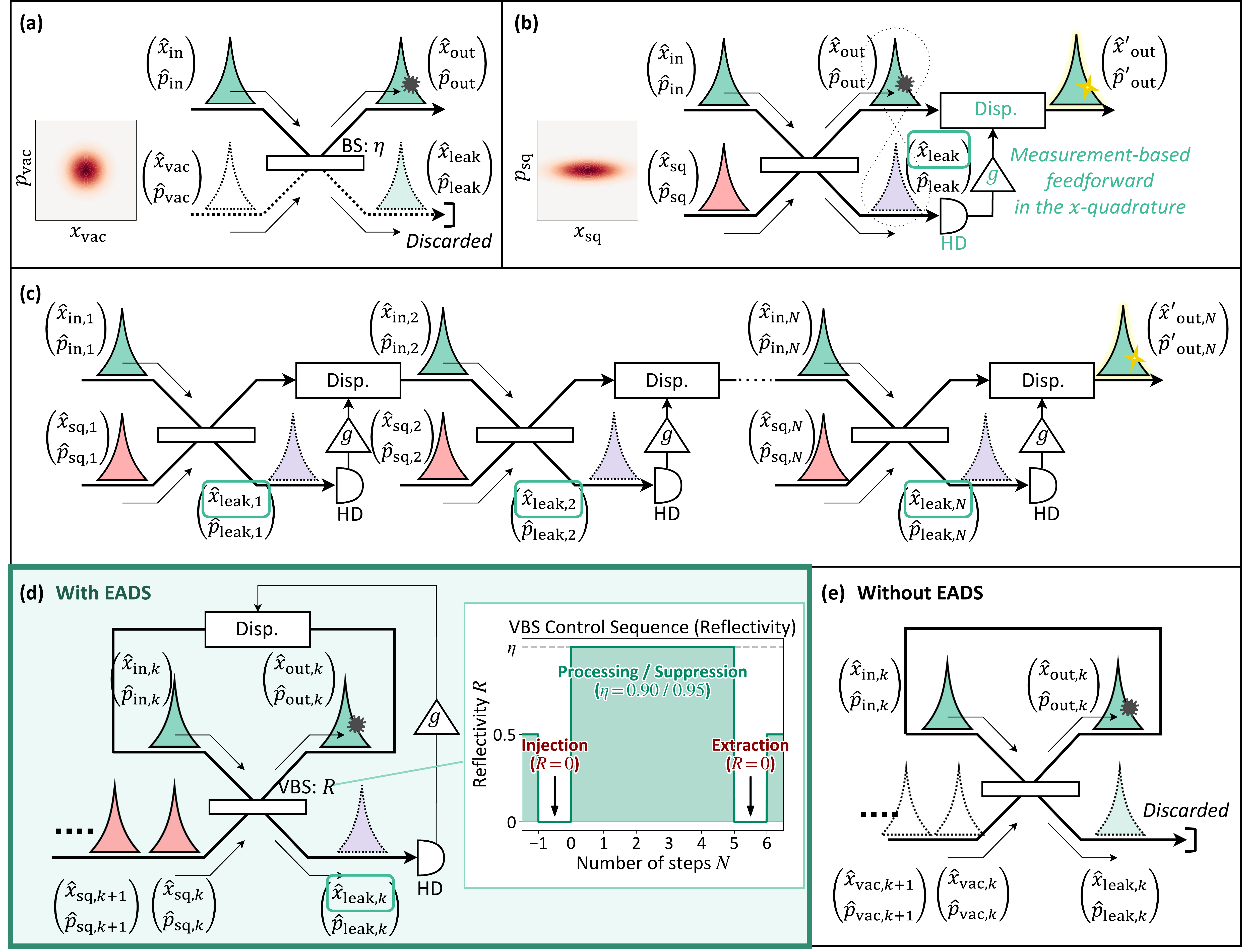}
    \caption{Conceptual schematic of the proposed EADS scheme.
    (a) Model of an optical loss channel. The target state interacts with an environmental vacuum state through a loss represented by the beam splitter (BS) with a transmittivity $\eta$, which leads to decoherence by quantum noise penalty in both quadratures.
    (b) Single-step suppression scheme. The environmental vacuum input is replaced by a $p$-squeezed vacuum state to reduce quantum noise penalty in one quadrature. The $x$-quadrature of the leaked mode is measured using a homodyne detector (HD). Based on the measurement result, a displacement operation (Disp.) is applied to the target mode via feedforward with gain $g$ to cancel the quantum noise penalty in the complementary quadrature.
    (c) Multi-step suppression. A cascaded implementation of (b) where the target state sequentially undergoes loss and suppression steps.
    (d) Loop-based implementation of multi-step suppression. Ancillary squeezed vacuum states are sequentially injected into the loop. The variable beam splitter (VBS) control sequence is designed to trap the target state in the loop, sustain it during the EADS cycles, and subsequently release it for measurement.
    (e) Loop-based implementation without EADS. The ancillary inputs are blocked, allowing environmental vacuum states to enter the system. The leaked modes are discarded, representing a decoherence process with an accumulated quantum-noise penalty in both quadratures. }
\label{fig:concept}
\end{figure*}

When a quantum system couples to an environment, decoherence inevitably occurs~\cite{Zurek2003}. In optical systems, the dominant source of the decoherence is photon loss. As illustrated in Fig.~\ref{fig:concept}\textcolor{blue}{(a)}, the loss of $1-\eta$ is modeled by a beam splitter interaction at reflectivity $\eta$~\cite{leonhardt2010essential}. As the target quantum state propagates through one input port, vacuum noise from the environment enters through the other. This interaction in the Heisenberg picture is described by the input-output relations:
\begin{align}
    \hat{x}_\mathrm{out} &= \sqrt{\eta}\hat{x}_{\mathrm{in}} + \sqrt{1-\eta}\hat{x}_{\mathrm{vac}}, \label{eq:(1)}\\
    \hat{p}_\mathrm{out} &= \sqrt{\eta}\hat{p}_{\mathrm{in}} + \sqrt{1-\eta}\hat{p}_{\mathrm{vac}},\label{eq:(2)}\\
    \hat{x}_\mathrm{leak} &= \sqrt{1-\eta}\hat{x}_{\mathrm{in}} - \sqrt{\eta}\hat{x}_{\mathrm{vac}},\label{eq:(3)}\\
    \hat{p}_\mathrm{leak} &= \sqrt{1-\eta}\hat{p}_{\mathrm{in}} - \sqrt{\eta}\hat{p}_{\mathrm{vac}},\label{eq:(4)}
\end{align}
where ($\hat{x}_{\rm{in}}$, $\hat{p}_{\rm{in}}$) and ($\hat{x}_{\rm{vac}}$, $\hat{p}_{\rm{vac}}$) denote the quadratures of the target input and environmental vacuum, respectively. The interaction results in the environmental output quadratures $\hat{x}_{\rm{leak}}$ and $\hat{p}_{\rm{leak}}$, which represent the leaked modes. These modes are subsequently discarded, leading to the loss of a portion of the target information and the introduction of quantum noise penalty from the environment into $(\hat{x}_{\text{out}}, \hat{p}_{\text{out}})$.

Crucially, we can suppress this decoherence by actively controlling and measuring the environment~\cite{Marek2004, zhang2018quantum}, as shown in Fig.~\ref{fig:concept}\textcolor{blue}{(b)}. The intuition is as follows: Instead of the vacuum state $(\hat{x}_{\rm{vac}}, \hat{p}_{\rm{vac}})$, we inject a $p$-squeezed vacuum state defined by $(\hat{x}_{\rm{sq}}, \hat{p}_{\rm{sq}})=(e^{r_{\text{a}}}\hat{x}_{\rm{vac}}, e^{-r_{\text{a}}}\hat{p}_{\rm{vac}})$ into the environmental port, where $r_{\text{a}}$ is the squeezing parameter of this state. This substitution reduces the quantum noise added to the $p$-quadrature of the target mode, while increasing the noise added to the $x$-quadrature. 
However, the $x$-quadrature information can be recovered using quantum correlations between the output and leaked modes. Specifically, the added noise in the $x$-quadrature is cancelled by measuring the $x$-quadrature of the leaked mode and applying a feedforward displacement to the target.

Mathematically, this process is equivalent to a measurement-induced squeezing gate~\cite{Filip2005}. From Eqs.~(\ref{eq:(1)}-\ref{eq:(4)}), by substituting the environmental vacuum quadratures $(\hat{x}_{\mathrm{vac}}, \hat{p}_{\mathrm{vac}})$ with the squeezed vacuum quadratures $(\hat{x}_{\mathrm{sq}}, \hat{p}_{\mathrm{sq}}) = (e^{r_{\text{a}}}\hat{x}_{\mathrm{vac}}, e^{-r_{\text{a}}}\hat{p}_{\mathrm{vac}})$ and feedforwarding the measurement outcome of $\hat{x}_{\mathrm{leak}}$ with a gain $g = \sqrt{(1-\eta)/\eta}$, the decoherence-\textit{suppressed} output quadratures ($\hat{x}'_{\rm{out}}$, $\hat{p}'_{\rm{out}}$) are given by:
\begin{align}
    \hat{x}'_{\mathrm{out}} &= \hat{x}_{\mathrm{out}}+\sqrt{\frac{1-\eta}{\eta}}\hat{x}_{\mathrm{leak}}=\frac{1}{\sqrt{\eta}}\hat{x}_{\mathrm{in}} \label{eq:feedfoward1},  \\
    \hat{p}'_{\mathrm{out}} &= \hat{p}_{\mathrm{out}}=\sqrt{\eta}\hat{p}_{\mathrm{in}}+\sqrt{1-\eta}e^{-r_{\text{a}}}\hat{p}_{\mathrm{vac}}. \label{eq:feedfoward2}
\end{align}
The noise in the $p$-quadrature vanishes in the limit of infinite squeezing ($r_{\text{a}} \to \infty$). This means that, as shown in Eqs.~(\ref{eq:feedfoward1}) and~(\ref{eq:feedfoward2}), the noise is eliminated from both quadratures, resulting in the complete suppression of decoherence. Note that this protocol squeezes the target state by a factor of $\sqrt{\eta}$ in the $p$-direction. Since this is a unitary effect, it represents a reversible deformation rather than decoherence; thus, a subsequent anti-squeezing operation allows for the full recovery of the original state if needed.

\begin{figure*}[t]
    \centering
    \includegraphics[width=0.75\linewidth]{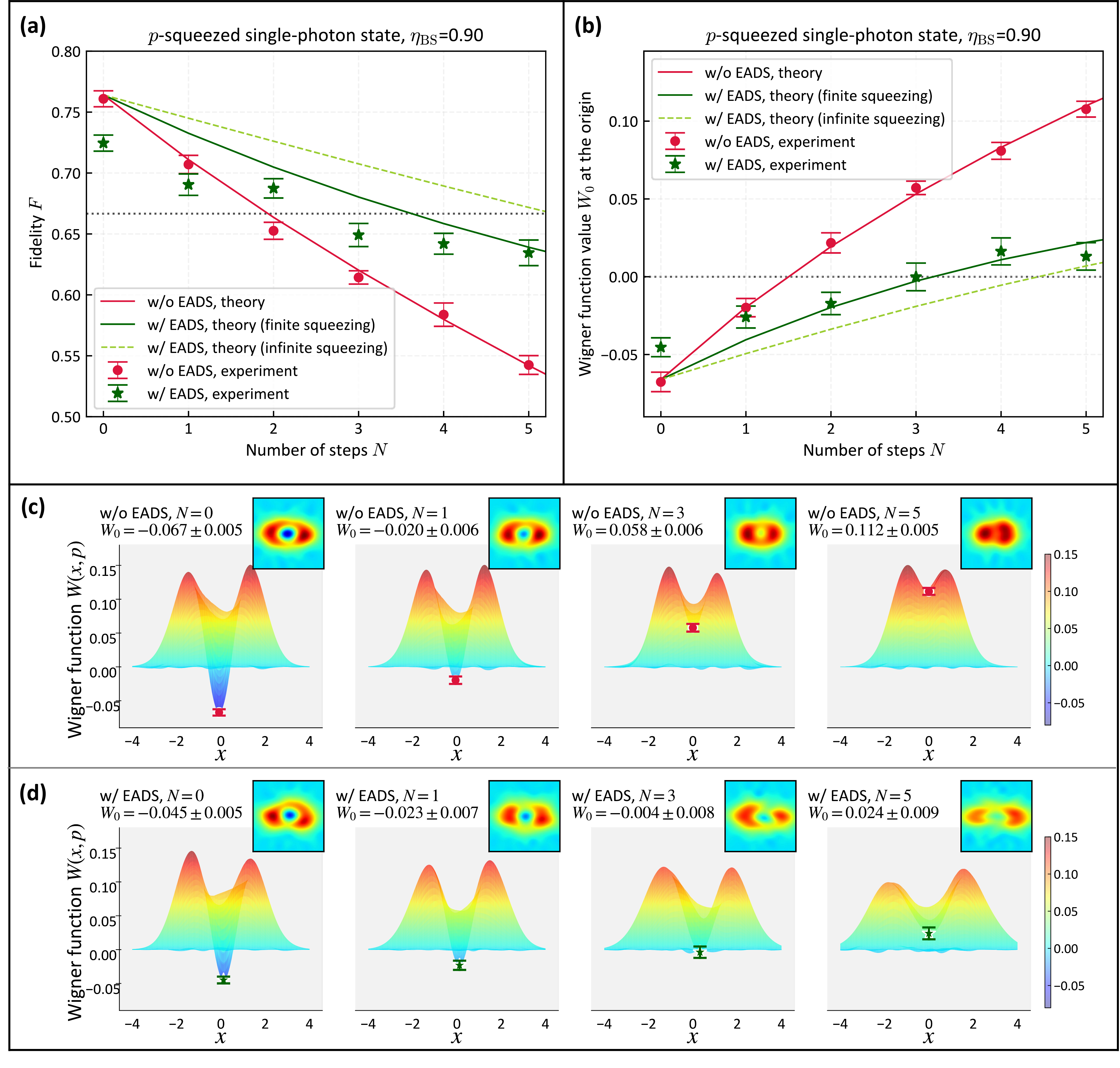}
    \caption{Experimental results for a $p$-squeezed single-photon state with 10\% beam splitter loss ($\hbar=1$). Error bars represent standard errors estimated using the bootstrap method. 
    (a) Decay of fidelity $F$ as a function of the number of steps. Red circles (w/o EADS) show the state degradation, while green stars (w/ EADS) maintain higher fidelity. 
    A horizontal black dotted line at $F=2/3$ indicates the no-cloning limit.
    (b) Decay of the Wigner function value at the origin, $W_0$, as a function of the number of steps. The \textit{suppressed} state (green) retains negative values longer than the \textit{unsuppressed} state (red).
    A horizontal black dotted line at $W_0=0$ marks the boundary below which essential quantum non-Gaussian features are preserved. In both (a) and (b), solid lines indicate the theoretical predictions based on experimental parameters and finite ancillary squeezing, and light-green dashed lines represent the theoretical limit assuming infinite ancillary squeezing (see Methods for details).
    (c, d) Wigner functions at steps $N=0, 1, 3, 5$. Panel (c) corresponds to the \textit{unsuppressed} case. Panel (d) corresponds to the \textit{suppressed} case, where the characteristic negative dip is more preserved. Insets show top views of the Wigner functions, where the horizontal and vertical axes represent the $x$ and $p$ quadratures, respectively.
    }
\label{fig:main_result_wigner}
\end{figure*}

\section{Experimental implementation\label{sec:setup}}

We experimentally verify this scheme by applying it to loss-sensitive non-Gaussian states with negative Wigner functions. To demonstrate a significant suppression effect, we simulate a process of repeated loss by applying the suppression in multiple steps.
A straightforward multi-step implementation in Fig.~\ref{fig:concept}\textcolor{blue}{(c)} would require cascading numerous beam splitters and detectors, resulting in a complex setup. To avoid this scaling issue, we perform time-domain multiplexed processing using a compact loop circuit in Fig.~\ref{fig:concept}\textcolor{blue}{(d)}. As the target state circulates within the loop, it experiences effective beam splitter loss at each round trip. We can inject squeezed light, measure the leak, and apply feedforward repeatedly in the time domain. This configuration not only demonstrates multi-step suppression but also simulates lifetime extension in a loop-based quantum memory~\cite{Marek2004, okuno2024time, kaneda2015time, simon2024experimental}.

We implement the experiment using our loop-based optical platform at 1545 nm~\cite{Yoshida2025}. The system consists of three main modules: a quantum light source, a loop circuit, and a homodyne detector.
The light source generates optical pulses at 61 ns intervals. Employing an optical switch, it can selectively provide either squeezed vacuum states (used as ancillary states) or non-Gaussian target states. These non-Gaussian states, squeezed single-photon or single-photon states, are created via photon subtraction~\cite{ourjoumtsev2006generating} from the squeezed vacuum output of an optical parametric amplifier.
The loop circuit includes a 61-ns optical delay line and a variable beam splitter (VBS) whose reflectivity can be switched every 61 ns. The loop phase is actively locked to maintain coherence between the circulating target and the injected ancilla.
Finally, the measurement basis of the homodyne detector is switchable every 61 ns, allowing us to sequentially measure both the leaked modes and the output states by appropriately controlling the measurement phase.
By performing measurements at various phases, we conduct quantum state tomography to evaluate the output state (see Methods for details).

The protocol proceeds as follows (see inset of Fig.~\ref{fig:concept}\textcolor{blue}{(d)}):
First, the VBS reflectivity is set from its default 50\% to 0\% to inject the target state in the loop.
Next, the system enters the decoherence suppression phase. The reflectivity is switched to $\eta$, simulating a loss of $1-\eta$ per round trip. 
For the \textit{suppressed} case shown in Fig.~\ref{fig:concept}\textcolor{blue}{(d)}, we inject $p$-squeezed vacuum states into the VBS, where they interfere with the target state at each step. During the suppression phase, the homodyne detector sequentially measures the $x$-quadrature of the leaked fields. After undergoing decoherence for $N$ round trips, the target state is released at the subsequent $(N+1)$-th step. Instead of the real-time optical feedforward depicted in Fig.~\ref{fig:concept}\textcolor{blue}{(d)}, we evaluate the $N$-step \textit{suppressed} state by numerically applying the feedforward operations in post-processing based on the sequence of leak measurement results. For comparison, we also implement the \textit{unsuppressed} case shown in Fig.~\ref{fig:concept}\textcolor{blue}{(e)}. In this configuration, by blocking the ancillary squeezed light, we repeatedly inject vacuum states into the loop. Consequently, the target state simply undergoes accumulated loss for $N$ round trips, and the final output state is evaluated by the homodyne detector at the subsequent $(N+1)$-th step. Our programmable setup allows us to vary the input state, the loss rate, and the number of steps, enabling a systematic comparison between the \textit{suppressed} and \textit{unsuppressed} cases within a single setup.
We note that this setup is designed to actively suppress the loss occurring at the VBS; however, the inherent internal propagation loss experienced during the loop round trip is left unsuppressed.

\section{Experimental results}

\begin{figure*}[t]
    \centering
    \includegraphics[width=0.90\linewidth]{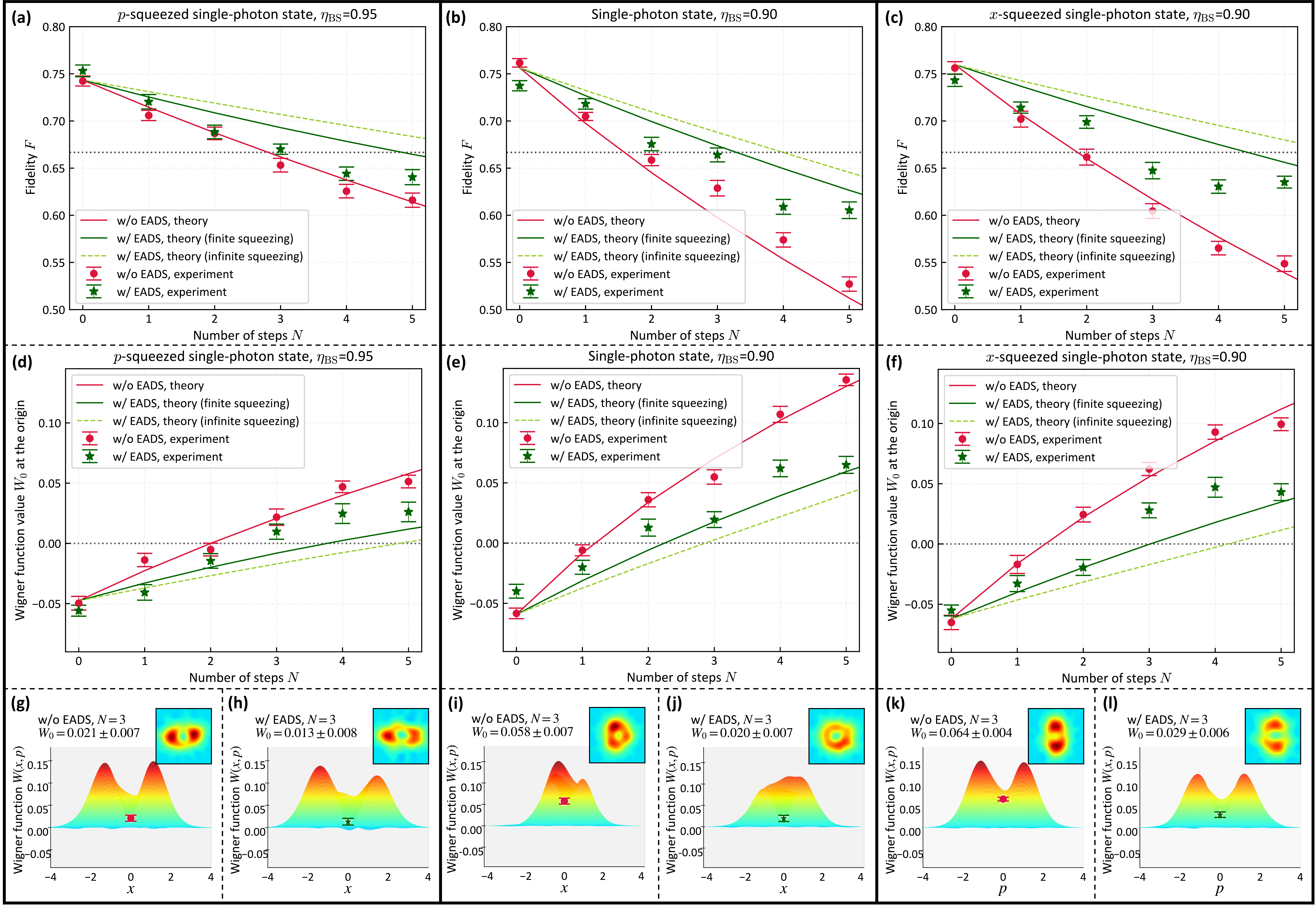}
    \caption{Experimental validation of decoherence suppression under varying conditions ($\hbar=1$). 
    The columns correspond to the following input states and beam splitter reflectivities. Left column (a, d, g, h): a $p$-squeezed single-photon state with 5\% loss ($\eta=0.95$). Middle column (b, e, i, j): a single-photon state with 10\% loss ($\eta=0.90$). Right column (c, f, k, l): an $x$-squeezed single-photon state with 10\% loss ($\eta=0.90$). 
    Error bars represent standard errors estimated using the bootstrap method.
    Panels (a–c) show the decay of fidelity $F$, and (d–f) show the decay of the Wigner function value at the origin, $W_0$, as a function of the step number $N$. Other definitions are the same as in Fig.~\ref{fig:main_result_wigner}.
    (g–l) Wigner functions at step $N=3$. Panels (g, i, k) correspond to the \textit{unsuppressed} cases, while (h, j, l) correspond to the \textit{suppressed} cases.}
\label{fig:comparison_wigner}
\end{figure*}

\begin{figure*}[htbp]
    \centering
    \includegraphics[width=0.90\linewidth]{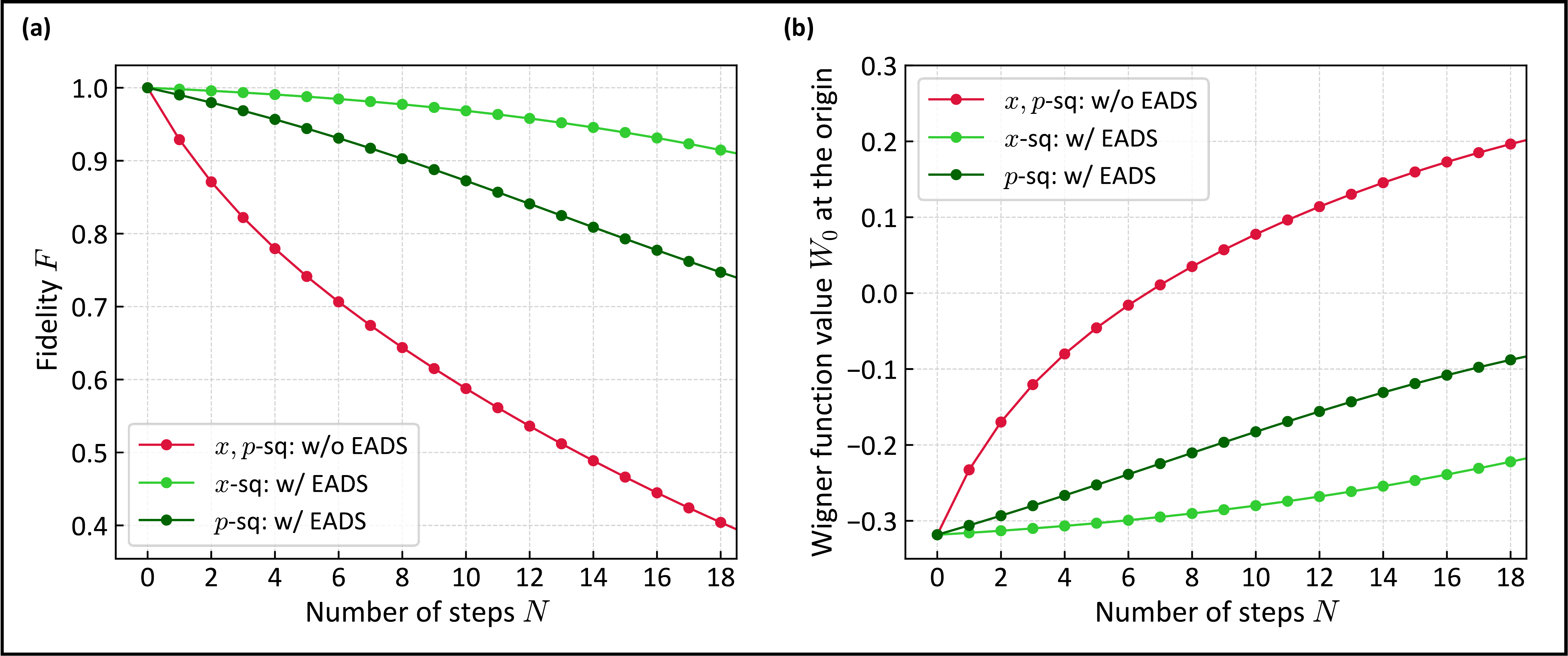}
    \caption{Numerical simulation of the dependence of EADS performance on the squeezed direction of the input state ($\hbar = 1$).
    The simulation assumes an ideal setup where both the loss experienced by the non-Gaussian state excluding the loop propagation and the round-trip loss accumulated within the loop are zero ($\eta_{\text{NG}} = \eta_{\text{loop}} = 1$). The beam splitter reflectivity is set to 90\%, and the ancillary squeezing level and its loss are set to the same values as those used for the theoretical curves in Figs.~\ref{fig:main_result_wigner} and \ref{fig:comparison_wigner}.
    Red curves represent the \textit{unsuppressed} case (where $x$- and $p$-squeezed states degrade identically). Dark green and light green curves represent the \textit{suppressed} cases for $p$-squeezed and $x$-squeezed input states, respectively.
    (a) Decay of fidelity $F$ and (b) decay of the Wigner function value at the origin, $W_0$, as a function of the step number $N$.}
    \label{fig:x_VS_p}
\end{figure*}

We perform experiments using three distinct non-Gaussian target states $\ket{\psi_{\mathrm{in}}}$ (single-photon state, $x$- or $p$-squeezed single-photon states) combined with two loss conditions (5\% and 10\% per step). These six scenarios serve to confirm the scheme's effectiveness across a range of experimental parameters.
We evaluate the performance by reconstructing the output density matrix $\hat{\rho}_{\mathrm{out}}$. We use the fidelity $F$ between $\ket{\psi_{\mathrm{in}}}$ and $\hat{\rho}_{\mathrm{out}}$ as a metric for state recovery and the Wigner function value at the origin, $W_0$, as an indicator of essential quantum non-Gaussian features~\cite{mari2012positive}. 
To benchmark these metrics, we introduce two physically meaningful thresholds. For fidelity, $F=2/3$ represents the no-cloning limit~\cite{grosshans2001quantum}; surpassing it demonstrates not only that the state's quantum characteristics are preserved beyond what any classical measure-and-prepare strategy could achieve, but also that no third party can possess an identical or better copy of the original quantum state. For the Wigner function, $W_0=0$ marks the definitive boundary separating quantum non-Gaussian states from classically simulatable states~\cite{mari2012positive}.
Note that $F$ is calculated after compensating for the deterministic squeezing effect described in Eqs.~(\ref{eq:feedfoward1}) and~(\ref{eq:feedfoward2}) (see Methods for details).

First, we show the result for a $p$-squeezed single-photon state under 10\% loss per step ($\eta=0.90$). Figure \ref{fig:main_result_wigner}\textcolor{blue}{(a)} shows the fidelity decay as a function of the number of steps. We observe that the \textit{unsuppressed} state (red circles) is progressively degraded by accumulated loss at each step. In contrast, the \textit{suppressed} state (green stars) shows a slower degradation of fidelity, remaining above the no-cloning limit for a larger number of steps. This confirms that our EADS scheme effectively mitigates the state degradation caused by loss. The slightly lower fidelity of the \textit{suppressed} case at $N=0, 1$ is due to the imperfect optical switch, which causes a small leakage of the ancillary squeezed light during the target state injection instead of the vacuum field. The experimental data for the \textit{suppressed} case follow the trend of the theoretical prediction (solid green line), which incorporates experimental imperfections such as finite ancillary squeezing and experimental losses. We also plot the theoretical limit assuming infinite ancillary squeezing (light-green dashed line). While this ideal case shows further improvement over the finite squeezing scenario (solid green line), the state still gradually degrades step by step. This is because our scheme addresses only the loss due to leakage at the VBS, leaving the inherent internal propagation loss within the loop unsuppressed ($\sim6\%$, see Table \ref{tab:measurement}).
To visualize the actual decay of the quantum states, we present the reconstructed Wigner functions in Figs.~\ref{fig:main_result_wigner}\textcolor{blue}{(c)} and \textcolor{blue}{(d)}. In the \textit{unsuppressed} case in Fig.~\ref{fig:main_result_wigner}\textcolor{blue}{(c)}, the characteristic non-Gaussian feature—specifically the negativity at the origin—decays due to photon loss. In contrast, the \textit{suppressed} case in Fig.~\ref{fig:main_result_wigner}\textcolor{blue}{(d)} retains this negative dip for a longer duration.
This preservation of quantum non-Gaussian features is quantified in Fig.~\ref{fig:main_result_wigner}\textcolor{blue}{(b)}, where the Wigner function value at the origin, $W_0$, is plotted as a function of the number of steps.
While the \textit{unsuppressed} state (red) rapidly loses its negativity, the \textit{suppressed} state (green) maintains a negative $W_0$ for more steps, demonstrating the effectiveness of our scheme.
Note that in Fig.~\ref{fig:main_result_wigner}\textcolor{blue}{(d)}, the distribution gradually becomes squeezed along the $p$-axis and elongated along the $x$-axis as the number of steps $N$ increases. This deformation arises from the deterministic squeezing operation applied in our protocol to counteract the loss as shown in Eqs.~(\ref{eq:feedfoward1}) and~(\ref{eq:feedfoward2}).

Figure \ref{fig:comparison_wigner} presents the results for three additional representative cases selected to provide a comprehensive comparison with Fig.~\ref{fig:main_result_wigner}: (i) A different loss rate: a $p$-squeezed single-photon state with 5\% loss per step.
(ii) A different input state: a standard single-photon state with 10\% loss per step.
(iii) A rotated input state: an $x$-squeezed single-photon state (rotated by $90^\circ$) with 10\% loss per step.
Across all these distinct regimes, the \textit{suppressed} cases (green) consistently exhibit superior performance in terms of both fidelity and Wigner function value at the origin compared to the \textit{unsuppressed} cases (red) in Figs.~\ref{fig:comparison_wigner}\textcolor{blue}{(a - f)}. 
This advantage is visually evident in the reconstructed Wigner functions at step $N=3$, shown in Figs.~\ref{fig:comparison_wigner}\textcolor{blue}{(g - l)}, where the \textit{suppressed} states in Figs.~\ref{fig:comparison_wigner}\textcolor{blue}{(h, j, l)} exhibit deeper dips compared to the \textit{unsuppressed} ones in Figs.~\ref{fig:comparison_wigner}\textcolor{blue}{(g, i, k)}.
We note that the \textit{suppressed} data in Figs.~\ref{fig:comparison_wigner}\textcolor{blue}{(a - f)} exhibit slight deviations from the theoretical predictions (solid lines). These fluctuations are somewhat larger than those in Figs.~\ref{fig:main_result_wigner}\textcolor{blue}{(a)} and \textcolor{blue}{(b)} but represent typical variances arising from technical imperfections in our current setup. Specifically, these imperfections are likely to arise from various factors, including phase-locking point offsets and fluctuations in the loop circuit and at interference points, as well as feedforward imperfections due to gain mismatch. Despite these technical noises, the qualitative advantage of the suppression scheme is robustly confirmed across all tested conditions (see Supplementary Section 2 for full experimental results).

In addition to the experimental results, our simulation shows that the EADS performance depends on the squeezed direction of the input state. To see this clearly, we focus on the ideal case with no loop loss and no initial input-state loss ($\eta_{\text{loop}}=\eta_{\text{NG}}=1$), as shown in Fig.~\ref{fig:x_VS_p}. 
We find that the $x$-squeezed single-photon state gives better fidelity $F$ and Wigner function values $W_0$ at the origin than the $p$-squeezed state. This happens mainly because the finite squeezing of the ancilla adds noise specifically in the $p$-direction, as seen in Eq.~(\ref{eq:feedfoward2}). For a $p$-squeezed input, this noise is added directly to the noise-sensitive squeezed part. In contrast, for an $x$-squeezed input, the noise affects the anti-squeezed part, which is less sensitive. 
Furthermore, the EADS operation itself squeezes the state in the $p$-direction. This makes an $x$-squeezed input rounder and more symmetric, while it stretches a $p$-squeezed input further, making it more fragile.
Although this difference is not clearly visible in our current experimental data due to the technical imperfections discussed above, these findings suggest a useful optimization. For known inputs, we can choose the suppression axis to minimize the noise. For unknown states, switching the direction (e.g., alternating $x$ and $p$) could help balance the effect and improve robustness.

\section{Discussion}

In this study, we have demonstrated a multi-step decoherence suppression scheme for optical non-Gaussian states. By actively controlling and monitoring the environment, we have achieved consistent improvements in fidelity and Wigner function value at the origin across diverse conditions.
Crucially, our loop-based implementation not only verified the multi-step decoherence suppression effect under
various conditions, but also serves as a proof-of-principle for extending the lifetime of quantum memories in the presence of loss. This capability is essential for buffering
probabilistic resource states and synchronizing timings in large-scale time-multiplexed architectures~\cite{makino2016synchronization, takeda2019demand, kaneda2015time, takeda2017universal, yoshikawa2016invited}.

The scheme demonstrated in this work is versatile and, in principle, applicable to arbitrary, unknown quantum states. While not applicable to purely dissipative losses such as absorption, our scheme addresses loss mechanisms in which the environmental modes can be physically controlled and monitored. In practical implementations, this approach enables the mitigation of losses originating from unwanted reflection or transmission at imperfect optical components, such as mirrors and lenses. Furthermore, in the presence of losses caused by imperfect spatial or temporal mode matching (e.g., coupling loss at interfaces or imperfect interference visibility between beams), by mode crosstalk (e.g., coupling between different spatial or polarization modes in fibers or waveguides), or by imperfect quantum transduction~\cite{zhang2018quantum, werner2026electro} (e.g., non-unit efficiency frequency conversion), the method remains applicable if the environmental modes can be filled with squeezed vacuum states and the mismatched components are subsequently measured.
Beyond optics, our scheme is potentially applicable to loss errors in other continuous-variable systems, such as microwave modes in superconducting circuits~\cite{campagne2020quantum, eickbusch2022fast} and vibrational modes in trapped ions~\cite{fluhmann2019encoding}. Furthermore, its compatibility with other error correction protocols and loss suppression schemes~\cite{chuang1997bosonic, gottesman2001encoding, eisert2004distillation, takahashi2010entanglement, LeJeannic2018} offers a promising route to reducing the overhead for fault tolerance~\cite{fukui2018high, guillaud2019repetition}. For instance, our scheme can be seamlessly integrated with quantum error correction codes, such as Gottesman-Kitaev-Preskill (GKP) states~\cite{gottesman2001encoding, campagne2020quantum}. Although GKP codes are robust against loss-induced errors (modeled as Gaussian random displacement noise), their fault-tolerant performance ultimately depends on keeping the effective loss rate below a specific threshold. By employing our EADS scheme to suppress the effective loss rate at the physical level, we can lower the threshold requirements for fault-tolerant quantum computing~\cite{fukui2018high, guillaud2019repetition}.

Thus, our work lays a foundation for high-fidelity, large-scale optical quantum processing, underpinning future advances in computing, communication, and sensing.

\section*{Methods}

\textbf{Experimental settings and parameters.}
We prepare three types of input states: $x$- or $p$-squeezed single-photon states, and single-photon states.
Target non-Gaussian states are generated via photon subtraction from a continuous-wave squeezed vacuum field. For the generation of $x$- or $p$-squeezed single-photon states, the initial squeezed vacuum has a measured squeezing level of 1.8 dB and an anti-squeezing level of 2.4 dB (corresponding to an estimated pure squeezing level of 3.5 dB). On the other hand, single-photon states are generated under weak squeezing conditions, with an estimated pure squeezing of 1.0 dB, thereby approximating a pure Fock state.
The ancillary squeezed state for decoherence suppression is $p$-squeezed, with a measured squeezing of 4.5 dB and anti-squeezing of 8.4 dB (corresponding to a pure squeezing of 9.7 dB).
For quantum state tomography, we perform homodyne measurements at 12 distinct phases, acquiring 1500 data points for each setting. Wigner functions are reconstructed from the homodyne data, and the fidelities are subsequently calculated. 
Theoretically, the feedforward gains are determined by the beam splitter reflectivity $\eta_{\text{BS}}$ and the internal loop loss $1-\eta_{\text{loop}}$ (see Supplementary Information for details). In the experiment, the gains are set based on the nominal reflectivities $\eta_{\text{BS}}$ = 0.90 or 0.95, together with the independently measured loop efficiency $\eta_{\text{loop}}$ = 0.94.

\vspace{2mm}
\textbf{Evaluation of output states.} 
To evaluate the performance of our scheme, we calculate the fidelity $F$ between the output state $\hat{\rho}_{\text{out}}$ and the target input state $\ket{\psi_{\text{in}}}$. 
As an inherent byproduct of the EADS scheme described in Eqs.~(\ref{eq:feedfoward1}) and~(\ref{eq:feedfoward2}), the output state undergoes a deterministic squeezing transformation. 
To isolate the non-Gaussian characteristics and assess the state's preservation, we compensate for this effect in post-processing by applying a numerical anti-squeezing operation, $\hat{S}(r_{\text{byp}})$, where $r_{\text{byp}}$ denotes the squeezing parameter of the deterministic byproduct ($r_{\text{byp}}=\frac{1}2{}\ln{\eta_{\text{BS}}}$). 
The compensated fidelity is thus defined as:
\begin{equation}
    F = \bra{\psi_{\text{in}}} \hat{S}(r_{\text{byp}}) \hat{\rho}_{\text{out}} \hat{S}^{\dagger}(r_{\text{byp}}) \ket{\psi_{\text{in}}}.
\end{equation}
To evaluate the peak negativity, $W_0$ is defined as the minimum value of the Wigner function when it attains negative values (not necessarily at the origin), and as its value at the origin $(x=p=0)$ otherwise.

\vspace{2mm}
\textbf{Theoretical model.}
Theoretical curves in Figs.~\ref{fig:main_result_wigner}, \ref{fig:comparison_wigner} and \ref{fig:x_VS_p} are derived from the model in Supplementary Information. The model is characterized by the following physical parameters: 
the squeezing parameter $r_{\rm{NG}}$ of the initial squeezed vacuum state used to generate the input non-Gaussian state; 
the loss $1-\eta_{\text{NG}}$ experienced by the non-Gaussian state excluding the loop propagation (including the effects of state-preparation, propagation, and measurement); 
the reflectivity $\eta_{\text{BS}}$ of the VBS; 
the round-trip loss $1-\eta_{\text{loop}}$ accumulated within the loop (excluding the VBS loss);
the squeezing parameter $r_{\text{a}}$ and the loss $1-\eta_{\text{a}}$ of the ancillary squeezed state. The actual experimental parameter values are summarized in Table~\ref{tab:measurement}.

\begin{table}[htbp]
  \centering
  \caption{Experimental parameters based on settings or measurement results.}
  \begin{tabularx}{0.49\textwidth}{X|c}
    \toprule
    \textbf{Parameter description (Symbol)} & \textbf{Value} \\
    \midrule
    Input squeezing level for squeezed single photons $(10\log_{10}e^{2r_{\rm{NG}}})$ & 3.5 dB \\
    Initial input-state efficiency $(\eta_{\text{NG}})$ & 0.62 \\
    VBS reflectivity $(\eta_{\text{BS}})$ & 0.90 or 0.95 \\
    Loop round-trip efficiency $(\eta_{\text{loop}})$ & 0.94 \\
    Ancillary squeezing level $(10\log_{10}e^{2r_{\rm{a}}})$ & 9.7 dB \\
    Ancilla efficiency $(\eta_{\text{a}})$ & 0.73 \\
    \bottomrule
  \end{tabularx}
  \label{tab:measurement}
\end{table}

While the parameters listed in Table~\ref{tab:measurement} represent typical values, actual experimental conditions are subject to slight fluctuations between measurements. Therefore, to generate the theoretical curves in Figs.~\ref{fig:main_result_wigner} and \ref{fig:comparison_wigner}, we determine the specific values of $\eta_{\text{NG}}$ and $\eta_{\text{loop}}$ for each dataset by fitting the experimental data of $W_0$ obtained in the \textit{unsuppressed} case. In this procedure, $\eta_{\text{NG}}$ and $\eta_{\text{loop}}$ are treated as free parameters and optimized independently for each experimental condition. Subsequently, these best-fit values ($\eta_{\text{NG}}^{\text{fit}}$, $\eta_{\text{loop}}^{\text{fit}}$) are employed to generate theoretical curves for $W_0$ in the \textit{suppressed} case and for $F$.
Table~\ref{tab:fitting_params} summarizes the fitting parameters for each experimental condition presented in this paper.

\begin{table}[h]
  \centering
  \caption{Fitting parameters.}
  \begin{tabularx}{0.49\textwidth}{X|c|c}
    \toprule
    \textbf{Target state and leakage condition} & $\boldsymbol{\eta^{\text{fit}}_{\text{NG}}}$ & $\boldsymbol{\eta^{\text{fit}}_{\text{loop}}}$ \\
    \midrule
    $p$-squeezed single-photon, 10\% loss (Figs.~\ref{fig:main_result_wigner}\textcolor{blue}{(a, b)})& 0.64 & 0.95 \\
    $p$-squeezed single-photon, 5\% loss (Figs.~\ref{fig:comparison_wigner}\textcolor{blue}{(a, d)})& 0.60 & 0.97 \\
    Single-photon state, 10\% loss (Figs.~\ref{fig:comparison_wigner}\textcolor{blue}{(b, e)})& 0.60 & 0.97 \\
    $x$-squeezed single-photon, 10\% loss (Figs.~\ref{fig:comparison_wigner}\textcolor{blue}{(c, f)})& 0.64 & 0.95 \\
    \bottomrule
  \end{tabularx}
  \label{tab:fitting_params}
\end{table}

\section*{Acknowledgements}
This work was partly supported by the Japan Science and Technology Agency (JST) Grants No. JPMJCR25I4, No. JPMJFR223R, No. JPMJMS2064, No. JPMJMS256I, and No. JPMJPF2221, and the Japan Society for the Promotion of Science (JSPS) KAKENHI Grants No. 23H01102, No. 23K17300, No. 24KJ0934, and No. 25KJ0782. A.M., R.S., and K.A. acknowledge financial support from The Forefront Physics and Mathematics Program to Drive Transformation (FoPM), WINGS Program, the University of Tokyo.
P.M. acknowledges the project 25-17472S of the Czech Science Foundation, the European Union’s HORIZON Research and Innovation Actions under Grant Agreement no. 101080173 (CLUSTEC) and the project CZ.02.01.010022 0080004649 (QUEENTEC) of EU and the Czech Ministry of Education, Youth and Sport. R.F. acknowledges the project No. 21-13265X of the Czech Science Foundation.
We thank Takato Yoshida and Hiroko Tomoda for their assistance with the experimental setup and providing valuable information.

\section*{Author contributions}
S.T. conceived and supervised the project. P.M. and R.F. proposed the methodology for data analysis and state evaluation. A.M. performed the experiments and analyzed the data with assistance from R.S., D.O., N.A. and K.A. T.K. and T.U. developed and provided the PPLN waveguide module. S.M., M.Y., and H.T. developed and provided the SSPD system. A.M. wrote the manuscript with input from all authors.

\section*{Competing interests}
The authors declare no competing interests.

\section*{Data availability}
The data that support the findings of this study are openly available in Zenodo at 
\url{https://zenodo.org/records/19449883}.

\section*{Code availability}
The code that supports the findings of this study is openly available in Zenodo at  \url{https://zenodo.org/records/19449883}.

%

\setupSupplementary 

\begin{center}
    {\Large \textbf{Supplementary Information}}
\end{center}

\maketitle

\section{Theoretical Model for Environment-Assisted Decoherence Suppression (EADS)}
In this note, we provide the complete derivation of the Gaussian channel model used to describe the multi-step EADS process. The primary objectives of this theoretical framework are to determine the appropriate feedforward gains required for the experimental implementation and to provide a unified model for calculating the theoretical curves of both the fidelity $F$ and the Wigner function values $W_0$ at the origin presented in the main text. To this end, we explicitly derive the recursive relations for the quadratures and present the analytical derivation of the output Wigner function.

\begin{figure}[!b]
    \label{fig:multi-step}
    \centering
    \includegraphics[width=0.5\linewidth]{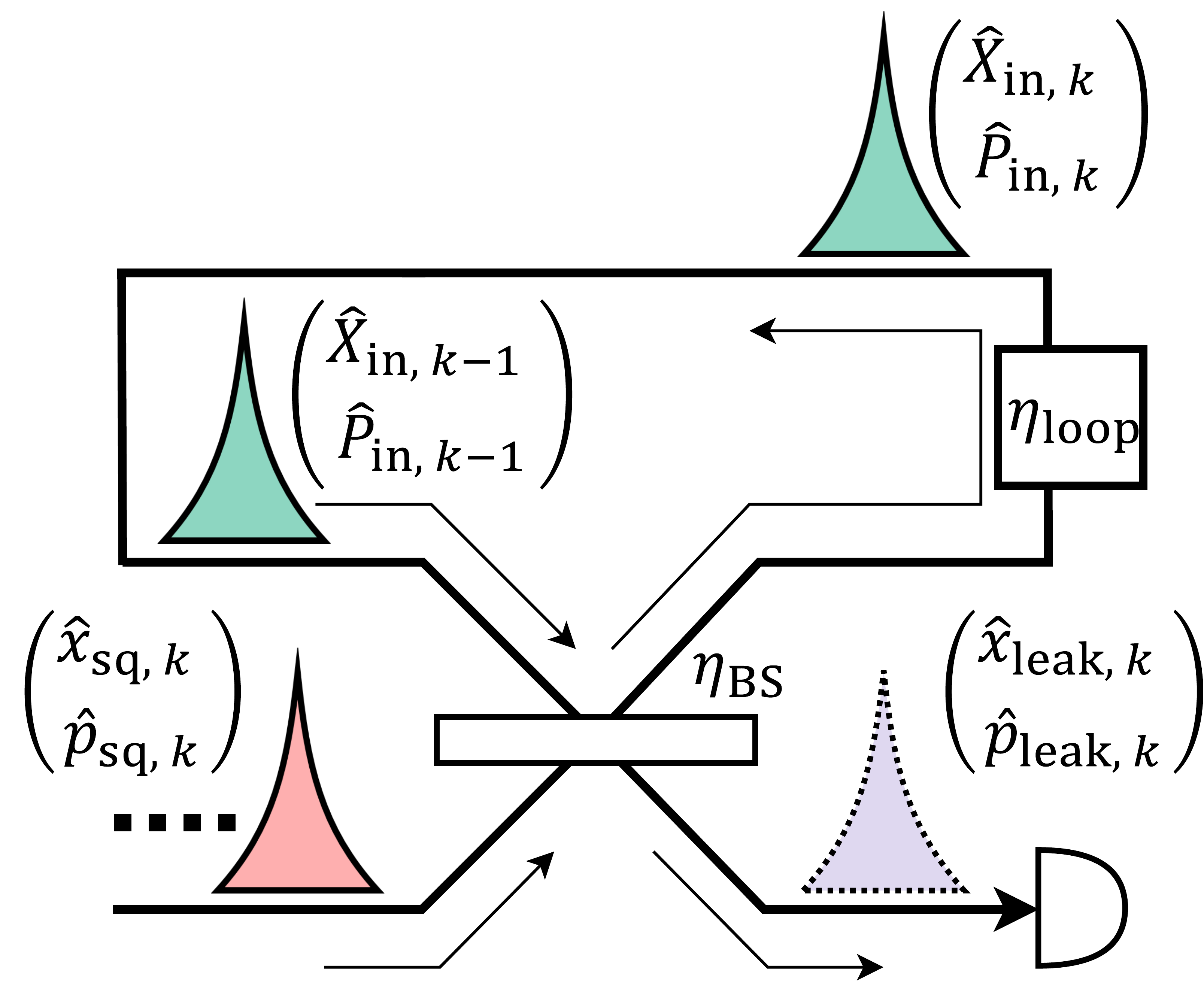}
    \caption{
    Schematic of the theoretical model for $N$-step EADS. 
    The diagram illustrates the $k$-th interaction step within the optical loop. The target state with quadratures $(\hat{X}_{\text{in}, k-1}, \hat{P}_{\text{in}, k-1})$ interacts with an ancillary $p$-squeezed vacuum state $(\hat{x}_{\text{sq}, k}, \hat{p}_{\text{sq}, k})$ at a beam splitter with reflectivity $\eta_{\text{BS}}$. The leaked quadratures are denoted as $(\hat{x}_{\text{leak}, k}, \hat{p}_{\text{leak}, k})$, and $\hat{x}_{\text{leak}, k}$ is measured via homodyne detection to provide the feedforward information required for EADS. After the interaction and accounting for the loop efficiency $\eta_{\text{loop}}$, the target state is transformed into $(\hat{X}_{\text{in}, k}, \hat{P}_{\text{in}, k})$ and recycled for the subsequent step.
    }
\label{fig:sup_setup}
\end{figure}

We model the system using the Heisenberg picture based on the optical circuit shown in Fig.~\ref{fig:sup_setup}. Let $(\hat{X}_{\text{in}, k-1}, \hat{P}_{\text{in}, k-1})$ be the quadrature operators of the target state inside the loop after $k-1$ round trips, where $(\hat{X}_{\text{in}, 0}, \hat{P}_{\text{in}, 0})$ represents the initial target state before entering the loop. At each step $ k $, an ancillary $p$-squeezed vacuum state with quadratures $(\hat{x}_{\text{sq}, k}, \hat{p}_{\text{sq}, k})$ is injected via a variable beam splitter (VBS) with reflectivity $\eta_{\text{BS}}$. 
The quadratures leaking from the VBS are denoted as $(\hat{x}_{\text{leak},k}, \hat{p}_{\text{leak},k})$, and we measure $\hat{x}_{\text{leak},k}$ using a homodyne detector to cancel out the anti-squeezed noise added to the target state at the VBS. To simplify the model, we assume that all propagation and measurement losses, except for the internal loop loss $1-\eta_{\text{loop}}$, are effectively incorporated into the state preparation stage of both the target and ancillary states. Therefore, we assume that $(\hat{X}_{\text{in},0}, \hat{P}_{\text{in},0})$ and $(\hat{x}_{\text{sq},k}, \hat{p}_{\text{sq},k})$ already include the effects of propagation and measurement losses, and that the homodyne measurement performed after the VBS is lossless. 
In terms of the experimental parameters defined in Table 1 of the main text, the losses incorporated into the initial target state $(\hat{X}_{\text{in},0},\hat{P}_{\text{in},0})$ and the ancillary state $(\hat{x}_{\text{sq},k},\hat{p}_{\text{sq},k})$ correspond to $1-\eta_{\text{NG}}$ and $1-\eta_{\text{a}}$, respectively.

The interaction at the VBS and the subsequent propagation/measurement are described by the following evolution. Based on the optical circuit shown in Fig.~\ref{fig:sup_setup}, the target quadratures $(\hat{X}_{\text{in}, k}, \hat{P}_{\text{in}, k})$ and the leaked quadratures $(\hat{x}_{\text{leak}, k}, \hat{p}_{\text{leak}, k})$ at each step satisfy the following relations:
\begin{align}
    \hat{X}_{\text{in}, k} &= \sqrt{\eta_{\text{loop}}}(\sqrt{\eta_{\text{BS}}} \hat{X}_{\text{in}, k-1} + \sqrt{1-\eta_{\text{BS}}} \hat{x}_{\text{sq}, k}) + \sqrt{1-\eta_{\text{loop}}}\hat{v}_{\text{loop},k}^x, \label{eq:Xk_init} \\
    \hat{P}_{\text{in}, k} &= \sqrt{\eta_{\text{loop}}}(\sqrt{\eta_{\text{BS}}} \hat{P}_{\text{in}, k-1} + \sqrt{1-\eta_{\text{BS}}} \hat{p}_{\text{sq}, k}) + \sqrt{1-\eta_{\text{loop}}}\hat{v}_{\text{loop},k}^p, \label{eq:Pk_init} \\
    \hat{x}_{\text{leak}, k} &= \sqrt{1-\eta_{\text{BS}}} \hat{X}_{\text{in}, k-1} - \sqrt{\eta_{\text{BS}}} \hat{x}_{\text{sq}, k}, \label{eq:xk_meas} \\
    \hat{p}_{\text{leak}, k} &= \sqrt{1-\eta_{\text{BS}}} \hat{P}_{\text{in}, k-1} - \sqrt{\eta_{\text{BS}}} \hat{p}_{\text{sq}, k}, \label{eq:pk_meas}
\end{align}
where $\hat{v}_{\text{loop},k}^{x,p}$ represents the vacuum noise operators associated with the internal loop loss.

As a first step, we focus on expressing $\hat{X}_{\text{in}, k}$ in terms of the measurable quadrature $\hat{x}_{\text{leak}, k}$ instead of the inaccessible ancillary quadrature $\hat{x}_{\text{sq}, k}$. By rearranging Eq.~\eqref{eq:xk_meas}, we can solve for $\hat{x}_{\text{sq}, k}$ as follows:
\begin{equation}
    \hat{x}_{\text{sq}, k} = \sqrt{\frac{1-\eta_{\text{BS}}}{\eta_{\text{BS}}}} \hat{X}_{\text{in}, k-1} - \frac{1}{\sqrt{\eta_{\text{BS}}}}\hat{x}_{\text{leak}, k}. \label{eq:xk_sub}
\end{equation}
Next, we substitute Eq.~\eqref{eq:xk_sub} into Eq.~\eqref{eq:Xk_init}:
\begin{align}
    \hat{X}_{\text{in}, k} &= \sqrt{\eta_{\text{loop}}}\left[ \sqrt{\eta_{\text{BS}}} \hat{X}_{\text{in}, k-1} + \sqrt{1-\eta_{\text{BS}}} \left( \sqrt{\frac{1-\eta_{\text{BS}}}{\eta_{\text{BS}}}} \hat{X}_{\text{in}, k-1} - \frac{1}{\sqrt{\eta_{\text{BS}}}}\hat{x}_{\text{leak}, k}\right) \right] \nonumber \\
    &\quad + \sqrt{1-\eta_{\text{loop}}}\hat{v}_{\text{loop},k}^x.
\end{align}
Using the relation $\sqrt{\eta_{\text{BS}}} + \frac{1-\eta_{\text{BS}}}{\sqrt{\eta_{\text{BS}}}} = \frac{1}{\sqrt{\eta_{\text{BS}}}}$, we obtain the recursive relation for $\hat{X}_{\text{in}, k}$ conditioned on the measurement outcome $\hat{x}_{\text{leak}, k}$:
\begin{equation}
    \hat{X}_{\text{in}, k} = \frac{\sqrt{\eta_{\text{loop}}}}{\sqrt{\eta_{\text{BS}}}}\hat{X}_{\text{in}, k-1} - \sqrt{\eta_{\text{loop}}}\sqrt{\frac{1-\eta_{\text{BS}}}{\eta_{\text{BS}}}} \hat{x}_{\text{leak}, k} + \sqrt{1-\eta_{\text{loop}}}\hat{v}_{\text{loop},k}^x.
\end{equation}
We iterate this relation from $k=1$ to $N$. After $N$ steps, the accumulated relation for the $X$-quadrature is:
\begin{align}
    \hat{X}_{\text{in}, N} &= \left(\sqrt{\frac{\eta_{\text{loop}}}{\eta_{\text{BS}}}}\right)^N \hat{X}_{\text{in}, 0} - \sqrt{\eta_{\text{loop}}}\sqrt{\frac{1-\eta_{\text{BS}}}{\eta_{\text{BS}}}} \sum_{k=1}^N \left(\sqrt{\frac{\eta_{\text{loop}}}{\eta_{\text{BS}}}}\right)^{N-k} \hat{x}_{\text{leak}, k} \nonumber \\
    &\quad + \sum_{k=1}^N \left(\sqrt{\frac{\eta_{\text{loop}}}{\eta_{\text{BS}}}}\right)^{N-k} \sqrt{1-\eta_{\text{loop}}}\hat{v}_{\text{loop},k}^x . \label{eq:XN_accum}
\end{align}
By feedforwarding the measured quadratures $\hat{x}_{\text{leak},k}$ to the target quadratures with the appropriate gain $g_k$ for each $k$-th step, the corrected final quadrature $\hat{X}_{\text{out}}$ is given by:
\begin{align}
    \hat{X}_{\text{out}} &= \hat{X}_{\text{in}, N} + \sum_{k=1}^N g_k \hat{x}_{\text{leak}, k} \nonumber \\
    &= \left(\sqrt{\frac{\eta_{\text{loop}}}{\eta_{\text{BS}}}}\right)^N \hat{X}_{\text{in}, 0} + \sum_{k=1}^N \left(\sqrt{\frac{\eta_{\text{loop}}}{\eta_{\text{BS}}}}\right)^{N-k} \sqrt{1-\eta_{\text{loop}}}\hat{v}_{\text{loop},k}^x , \label{eq:X_final}
\end{align}
where the gain $g_k$ is identified as:
\begin{align}
    g_k = \sqrt{\eta_{\text{loop}}}\sqrt{\frac{1-\eta_{\text{BS}}}{\eta_{\text{BS}}}} \left(\sqrt{\frac{\eta_{\text{loop}}}{\eta_{\text{BS}}}}\right)^{N-k}. \label{eq:sup_gain_def}
\end{align}
Equation \eqref{eq:X_final} represents the final input-output relation for the $X$-quadrature, showing that the anti-squeezed noise terms $\hat{x}_{\text{sq},k}$ are successfully eliminated. In the actual experiment, the feedforward gains are set according to Eq.~(\ref{eq:sup_gain_def}) to achieve this noise cancellation.

Next, we consider the $P$-quadrature. Since no feedforward operation is applied to this component, the final $P$-quadrature $\hat{P}_{\text{out}}$ is simply the result of iterating the relation in Eq.~\eqref{eq:Pk_init} $N$ times:
\begin{align}
    \hat{P}_{\text{out}} &= \hat{P}_{\text{in}, N} \nonumber\\
    &= (\sqrt{\eta_{\text{loop}}\eta_{\text{BS}}})^N \hat{P}_{\text{in}, 0} + \sum_{k=1}^N (\sqrt{\eta_{\text{loop}}\eta_{\text{BS}}})^{N-k} \left( \sqrt{\eta_{\text{loop}}(1-\eta_{\text{BS}})} \hat{p}_{\text{sq}, k} + \sqrt{1-\eta_{\text{loop}}}\hat{v}_{\text{loop},k}^p \right). \label{eq:P_final}
\end{align}
This provides the final input-output relation for the $P$-quadrature.

The transformation from the initial quadratures $(\hat{X}_{\text{in}, 0}, \hat{P}_{\text{in}, 0})$ to the final quadratures $(\hat{X}_{\text{out}},$ $ \hat{P}_{\text{out}})$ in Eqs.~\eqref{eq:X_final} and \eqref{eq:P_final} can be described as a Gaussian channel. In general, a single-mode Gaussian channel is characterized by its action on the first and second quadrature moments. For an arbitrary single-mode state with mean vector $\bar{\bm{\mu}}$ and covariance matrix $V$, the channel is defined as~\cite{walschaers2021non}:
\begin{align}
    \bar{\bm{\mu}} &\mapsto S \bar{\bm{\mu}}+\bm{\mu_0}, \label{eq:map_mean} \\
    V &\mapsto S V S^T + V^{\text{c}}, \label{eq:map_cov}
\end{align}
where $S$ represents the scaling matrix and $V^{\text{c}}$ is the covariance matrix of the added noise. The vector $\bm{\mu}_0$ serves to displace the quantum state to a different location in phase space.

In our experiment, we assume an input state with zero mean ($\bar{\bm{\mu}}=\bm{0}$). Based on the input-output relations derived in Eqs.~\eqref{eq:X_final} and \eqref{eq:P_final}, the scaling matrix $S$ and the vector $\bm{\mu}_0$ are explicitly given by:
\begin{align}
    &S = \sqrt{ \eta_{\text{loop}}^N} 
    \begin{pmatrix} 
        \eta_{\text{BS}}^{-N/2} & 0 \\ 
        0 & \eta_{\text{BS}}^{N/2} 
    \end{pmatrix},
    \label{eq:sup_byp}\\
    &\bm{\mu}_0 = \bm{0}.
\end{align}
The covariance matrix of the added noise is diagonal and takes the form $V^{\text{c}} = \text{diag}(v_1, v_2)$ in our case. The diagonal elements $(v_1, v_2)$ are determined by summing the variance contributions of all independent noise terms in the final $X$- and $P$-quadratures. 
Assuming that the vacuum noise operators $\hat{v}_{\text{loop},k}^{x}$ at different steps are independent and satisfy $\langle \hat{v}_{\text{loop},k}^{x} \hat{v}_{\text{loop},k'}^{x} \rangle = \delta_{kk'}/2$, the $X$-quadrature noise variance $v_1$ is derived from Eq.~\eqref{eq:X_final} as:
\begin{equation}
    v_1 =  \frac{1-(\eta_{\text{loop}}/\eta_{\text{BS}})^N}{2(1-\eta_{\text{loop}}/\eta_{\text{BS}})} \left( 1-\eta_{\text{loop}} \right).
    \label{eq:sup_v1}
\end{equation}
Similarly, assuming that the operators at different steps are independent and satisfy $\langle \hat{v}_{\text{loop},k}^p \hat{v}_{\text{loop},k'}^p \rangle = \delta_{kk'}/2$ and $\langle \hat{p}_{\text{sq},k} \hat{p}_{\text{sq},k'} \rangle =  \left(\frac{\eta_{\text{a}}}{2} e^{-2r_{\text{a}}} + \frac{1-\eta_{\text{a}}}{2}\right)\delta_{kk'}$, the $P$-quadrature noise variance $v_2$ is derived from Eq.~\eqref{eq:P_final} as:
\begin{equation}
    v_2 = \frac{1-(\eta_{\text{loop}}\eta_{\text{BS}})^N}{2(1-\eta_{\text{loop}}\eta_{\text{BS}})} \left( \eta_{\text{loop}}(1-\eta_{\text{BS}})(\eta_{\text{a}} e^{-2r_{\text{a}}} + 1-\eta_{\text{a}}) + 1-\eta_{\text{loop}} \right),
    \label{eq:sup_v2}
\end{equation}
where $r_{\text{a}}$ and $\eta_{\text{a}}$ denote the squeezing parameter and the preparation efficiency of the ancilla states, respectively.

Given the Gaussian channel characterized by $S$ and $V^{c}$ as formulated above, the output Wigner function $W_{\text{out}}(\boldsymbol{x})$ with phase-space coordinates $\boldsymbol{x}=(X_{\text{out}}, P_{\text{out}})^T$ is known to be expressed as the convolution of the initial Wigner function $W_{\text{in}}$ and a Gaussian noise kernel \cite{walschaers2021non}:
\begin{equation}
    W_{\text{out}}(\boldsymbol{x}) = \frac{1}{2\pi\sqrt{\det V^{\text{c}}}} \int W_{\text{in}}(S^{-1}(\boldsymbol{x}-\boldsymbol{y})) \exp\left( -\frac{1}{2}\boldsymbol{y}^T (V^{\text{c}})^{-1} \boldsymbol{y} \right) \frac{d\boldsymbol{y}}{\det S},
\end{equation}
where $\boldsymbol{y}=(y_x, y_p)^T$ is the integration variable over the entire phase space. 

Based on the above theoretical model, the output quantum state after implementation of the EADS scheme for an arbitrary input state can be calculated. In this work, we employed this model to derive the theoretical curves for the fidelity $F$ and the Wigner function value $W_0$ at the origin.

\vspace{10mm}

\section{Comprehensive Experimental Results}
In the demonstration of the EADS, we acquire experimental data under six different conditions, consisting of three types of input states and two loss rates. Figures 2 and 3 in the main text present results from four of these conditions, showing only a representative subset of the data. In this section, we provide the complete set of experimental results for all six conditions, including the evolution of the fidelity, the Wigner function values at the origin, and the reconstructed Wigner functions for all steps.

\begin{figure}[H]
  \centering
  \includegraphics[width=0.95\linewidth]{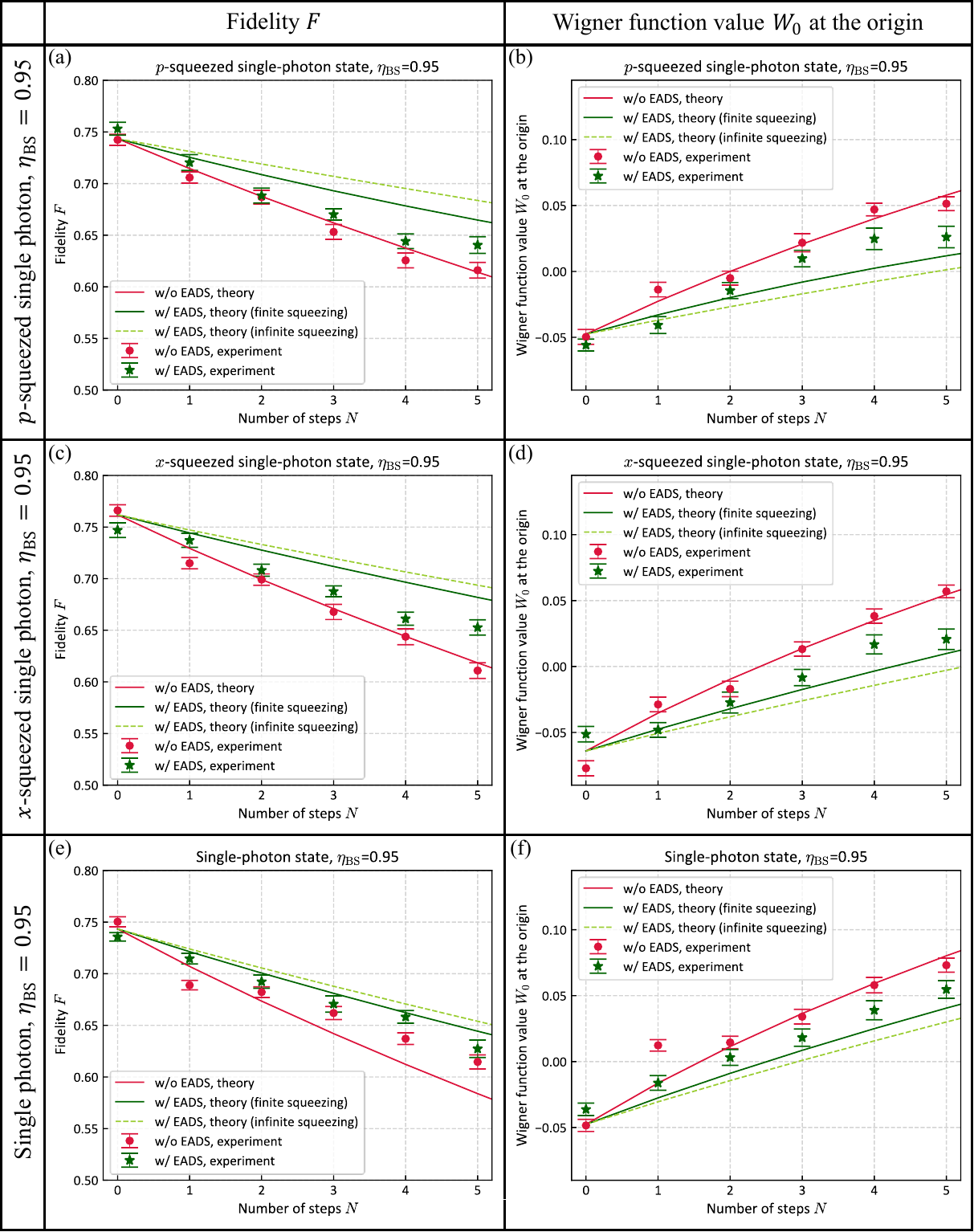}
  \caption{
Experimental results for $N$-step decoherence suppression with 5\% beam splitter loss ($\eta_{\text{BS}} = 0.95$).
The results are shown for three different input states: (a, b) $p$-squeezed single-photon state, (c, d) $x$-squeezed single-photon state, and (e, f) single-photon state. Red circles show the unsuppressed case. Green stars show the suppressed case. Solid lines indicate theoretical predictions based on experimental parameters. Light-green dashed lines represent the theoretical limit assuming infinite ancillary squeezing ($r_{\text{a}} \to \infty, \eta_{\text{a}} = 1.0$).
Panels (a, c, e) show the decay of fidelity $F$, and panels (b, d, f) show the evolution of the Wigner function value $W_0$ at the origin, which strictly corresponds to the minimum value when $W_0 < 0$ and the value at the origin when $W_0 \geq 0$. Error bars represent standard errors estimated using the bootstrap method.}
      \label{fig:sup_fig1}
\end{figure}

\begin{figure}[H]
  \centering
  \includegraphics[width=0.95\linewidth]{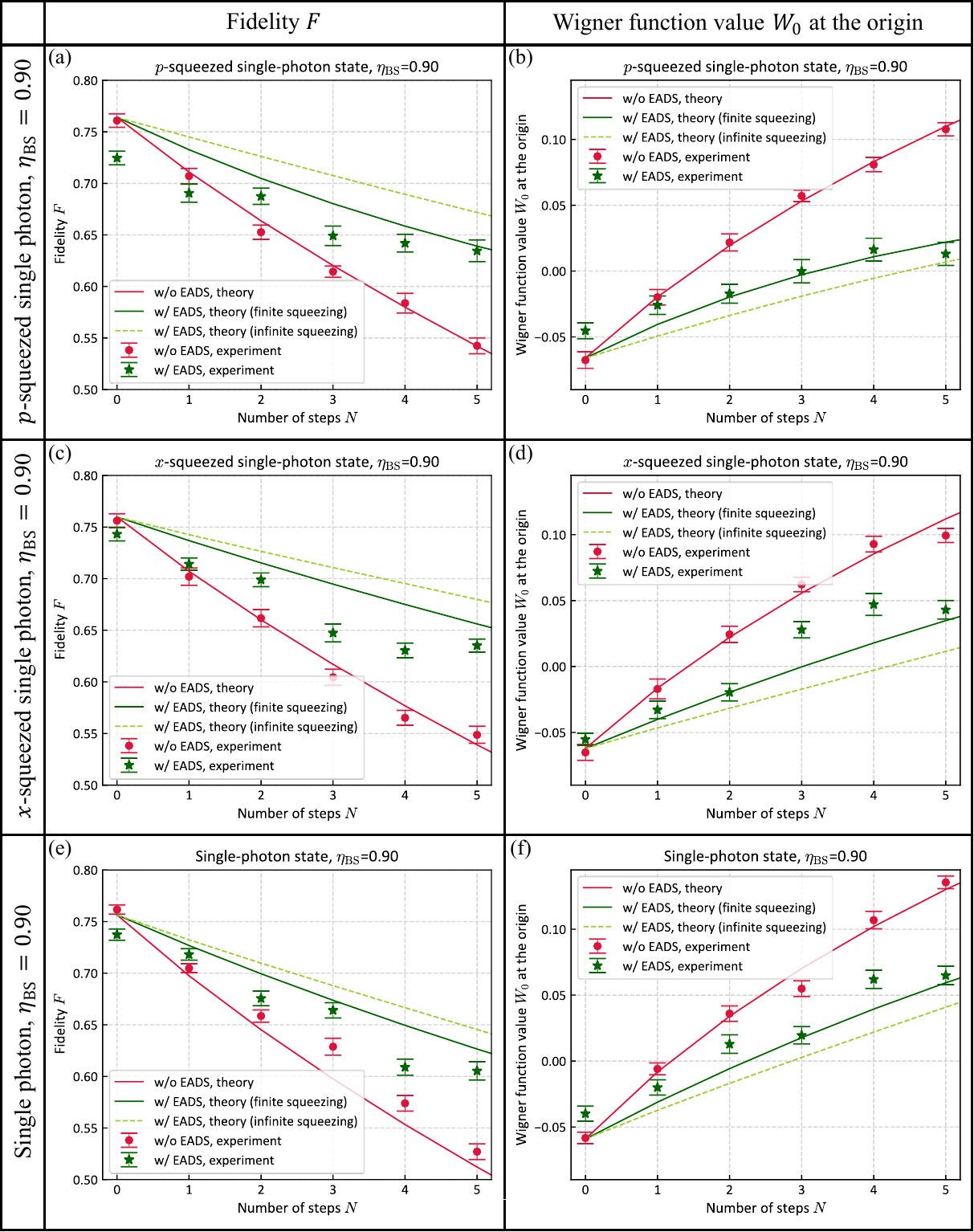}
  \caption{
    Experimental results for $N$-step decoherence suppression with 10\% beam splitter loss ($\eta_{\text{BS}} = 0.90$).
    All conditions (except for $\eta_{\text{BS}}$) and the plot notations are the same as in Fig.~\ref{fig:sup_fig1}.}
    \label{fig:sup_fig2}
\end{figure}

\begin{figure}[H]
  \centering
  \includegraphics[width=0.95\linewidth]{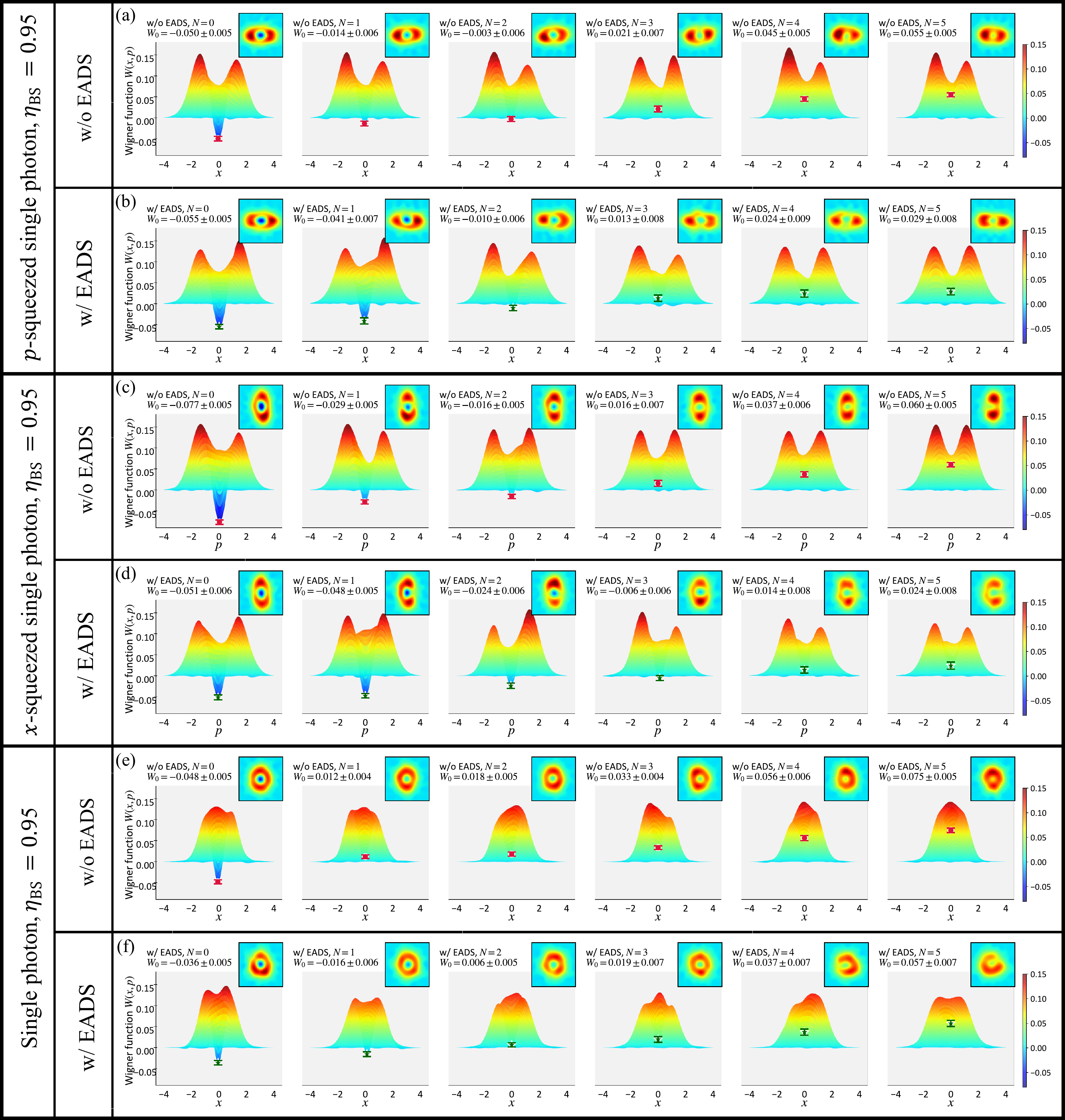}
    \caption{Experimental results for Wigner functions with 5\% beam splitter loss ($\eta_{\text{BS}} = 0.95$).
    The results are shown for three different input states: (a, b) $p$-squeezed single-photon state, (c, d) $x$-squeezed single-photon state, and (e, f) single-photon state. In each panel, the upper row shows the unsuppressed case, and the lower row shows the suppressed case for interaction steps $N=0$ to 5. Insets show top views of the Wigner functions, where the horizontal and vertical axes represent the $x$ and $p$ quadratures, respectively. The value of $W_0$ shown at the top of each Wigner function plot corresponds to the minimum value when $W_0<0$ and the value at the origin when $W_0 \geq 0$, as indicated by the red or green marker in the plot.}
    \label{fig:sup_fig3}
\end{figure}

\begin{figure}[H]
  \centering
  \includegraphics[width=0.95\linewidth]{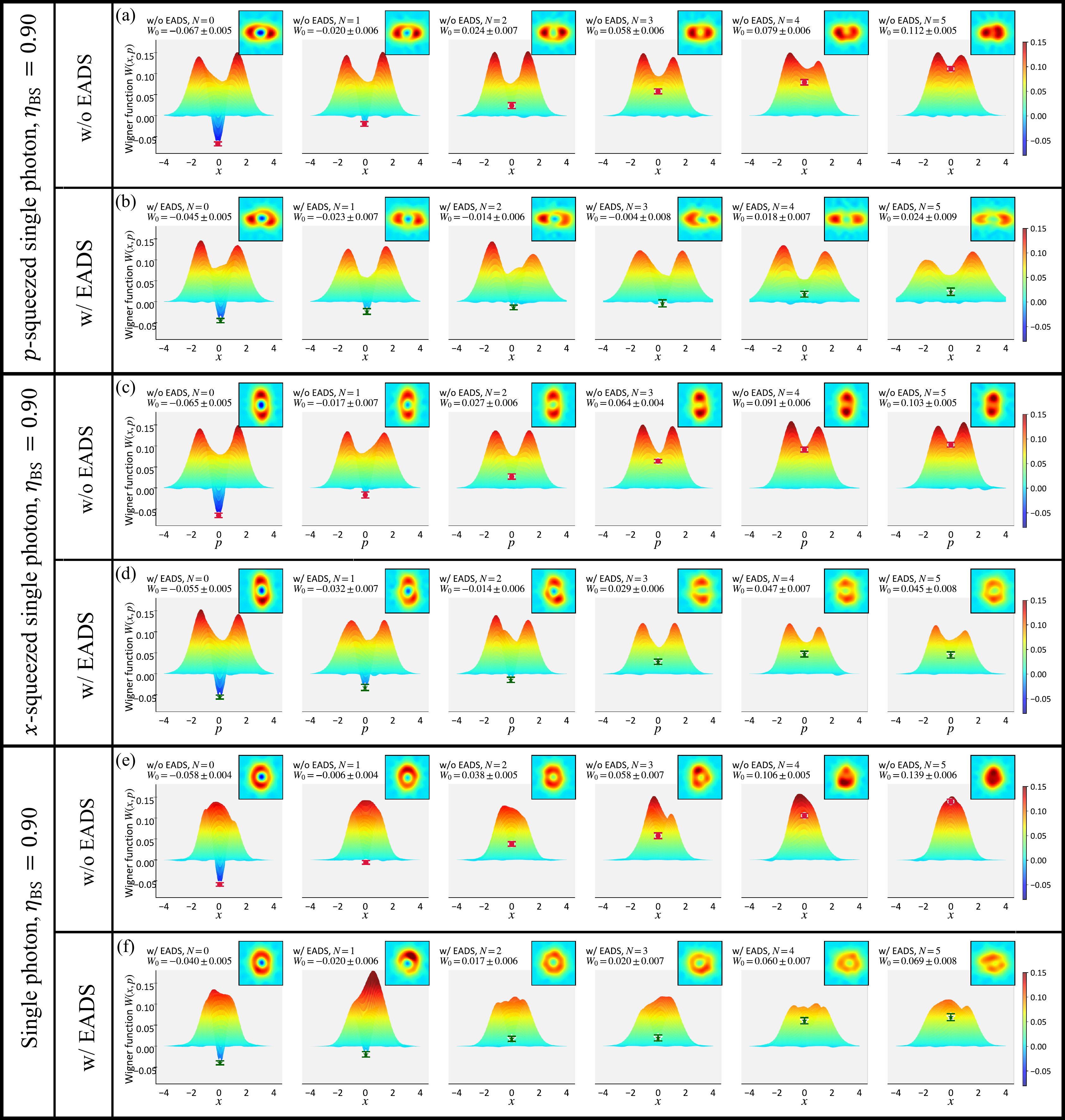}
  \caption{Experimental results for Wigner functions with 10\% beam splitter loss ($\eta_{\text{BS}} = 0.90$).
  All conditions (except for $\eta_{\text{BS}}$) and plot notations are the same as in Fig.~\ref{fig:sup_fig3}.}
  \label{fig:sup_fig4}
\end{figure}

\vspace{10mm}


%

\label{LastBibItem} 

\end{document}